\documentclass[11pt]{article}

\usepackage[letterpaper, margin=1in]{geometry}
\usepackage{amsthm,amsmath,amsfonts,amssymb}
\usepackage[authoryear]{natbib}
\usepackage{graphicx}
\usepackage{booktabs}
\usepackage{xcolor}
\usepackage{mathtools}
\usepackage[colorlinks,citecolor=blue,urlcolor=blue]{hyperref}

\newcommand{\bx}{\boldsymbol{x}}
\newcommand{\bs}{\boldsymbol{s}}
\newcommand{\bw}{\boldsymbol{w}}
\newcommand{\bv}{\boldsymbol{v}}
\newcommand{\bz}{\boldsymbol{z}}
\newcommand{\btheta}{\boldsymbol{\theta}}
\newcommand{\bepsilon}{\boldsymbol{\epsilon}}
\newcommand{\bzero}{\boldsymbol{0}}
\newcommand{\LF}{\text{LF}}
\newcommand{\HF}{\text{HF}}
\DeclareMathOperator{\vect}{vec}
\DeclareMathOperator{\cov}{cov}
\DeclareMathOperator{\blockDiag}{blockDiag}
\DeclareMathOperator{\logit}{logit}

\begin{document}

\title{A Multi-Fidelity Tensor Emulator for Spatiotemporal Outputs: Emulation of Arctic Sea-Ice Dynamics}
\author{
    Tristan Contant\thanks{Department of Statistics, Colorado State University. Email: \texttt{tristan.contant@colostate.edu}}
    \and Yawen Guan\thanks{Department of Statistics, Colorado State University. Email: \texttt{yawen.guan@colostate.edu}}
    \and Ander Wilson\thanks{Department of Statistics, Colorado State University. Email: \texttt{ander.wilson@colostate.edu}}
    \and Adrian K. Turner\thanks{Los Alamos National Laboratory. Email: \texttt{akt@lanl.gov}}
    \and Deborah Sulsky\thanks{Department of Mathematics and Statistics, University of New Mexico. Email: \texttt{sulsky@math.unm.edu}}
}
\date{}
\maketitle

\begin{abstract}
    Numerical models are widely used to simulate the Earth system, but they are computationally expensive and often depend on many uncertain input parameters.
    Calibration and uncertainty quantification require running them across many input configurations, so their effective use is costly.
    Statistical emulation provides a practical alternative for efficiently exploring model behavior.
    We are motivated by the Arctic sea-ice component of the Energy Exascale Earth System Model (MPAS-Seaice), which generates large spatiotemporal outputs at multiple spatial resolutions, with high-resolution (or high-fidelity, HF) simulations being more accurate but computationally more expensive than lower-resolution (low-fidelity, LF) simulations.
    Multi-fidelity (MF) emulation integrates information across resolutions to construct efficient and accurate surrogate models, yet existing approaches struggle to scale to large spatiotemporal data.
    We develop a MF emulator that combines tensor decomposition for dimensionality reduction, Gaussian process priors for flexible function approximation, and an additive discrepancy model to capture systematic differences between LF and HF data.
    The proposed framework scales to complex spatiotemporal fields while keeping predictions accurate and uncertainty well calibrated.
    In the MPAS-Seaice analysis, it attains the lowest prediction error of all emulators considered, including separable-covariance and neural-network approaches, especially throughout the summer melt season, when sea-ice concentration varies most across model settings and matters most scientifically, and best reproduces the spatiotemporal dependence of the output.
    By combining many inexpensive low-resolution runs with a few expensive high-resolution ones, the emulator substantially reduces the simulation cost of exploring model behavior, and suits large-scale studies involving complex physical models.
\end{abstract}


\section{Introduction} \label{sec:intro}

Numerical models are widely used to simulate the Earth system and study the complex physical processes that drive variability across interacting atmospheric, oceanic, terrestrial, and cryospheric components \citep{chen2021}.
Within this interconnected system, Arctic sea ice is a critical component that regulates ocean-atmosphere exchanges, influences large-scale circulation, and supports polar ecosystems, while also affecting human activities within coastal communities and marine navigation \citep{stro2018, meier2022}.
Accurately representing sea-ice dynamics in numerical models requires careful parameterization and tuning, as well as simulations at sufficiently high spatial resolution to resolve key physical and biogeochemical processes and localized phenomena.
However, high-resolution model simulations are computationally expensive, severely limiting the number of parameter combinations that can be explored \citep{schn2017}.
To address these challenges, statistical emulators offer efficient alternatives that enable broader exploration of parameter space and improved uncertainty quantification, while substantially reducing computational cost \citep{suda2022}.

MPAS-Seaice \citep{turn2022} is the sea-ice component of the Energy Exascale Earth System Model \citep[E3SM;][]{Leung2020}, the U.S.\ Department of Energy's Earth System Model.
The model is forced by atmospheric and oceanic fields, including winds, ocean currents, temperature, radiation, and sea surface conditions, that drive thermodynamic and dynamic evolution of the sea ice \citep{urre2016}.
MPAS-Seaice is a key modeling tool for advancing the scientific understanding of sea-ice dynamics and thermodynamics.
As part of the E3SM model, MPAS-Seaice supports a wide range of research efforts, from evaluating coupled Earth-system interactions to informing model development and assessment across the scientific community \citep{Leung2020, turn2022}.
For a given set of geophysical input parameters, MPAS-Seaice produces spatiotemporal outputs describing the evolution of the sea-ice state on an unstructured spherical centroidal Voronoi tessellation, provided by the Model for Prediction Across Scales \citep[MPAS;][]{Ringler2010} modeling framework.
For additional details regarding the MPAS-Seaice physics and the forcing information producing our model runs, see Supplementary Material~A.

MPAS-Seaice can be run at varying spatial resolutions.
Finer spatial resolution (henceforth high-fidelity or HF) captures small-scale processes like lead formation (openings in the ice) but is computationally expensive.
Coarse-resolution simulations (henceforth low-fidelity or LF) are cheaper and allow more runs, but they can miss important fine-scale dynamics.
Consequently, outputs across fidelities are not related by simple spatial interpolation and instead exhibit systematic differences in their representation of sea-ice processes.
In this work, we develop a multi-fidelity (MF) model, leveraging both LF and HF outputs, to emulate high-resolution MPAS-Seaice simulations for new inputs.

Emulation in this context is particularly challenging due to high-dimensional spatiotemporal outputs, a moderate number of design parameters, and a limited number of available simulations.
Specifically, we ran MPAS-Seaice at two resolutions: $n_s^{\LF} = 11{,}661$ spatial locations for the LF and $n_s^{\HF} = 33{,}530$ for the HF.
For both LF and HF simulations, the output consists of $132$ time points corresponding to monthly data over an 11-year period.
The runs varied by a set of $p = 6$ input design parameters, with $n_x^{\LF} = 89$ and $n_x^{\HF} = 21$ runs for the LF and HF configurations, respectively.
Hence, we are mapping a set of $6$ input parameters to high-dimensional spatiotemporal output based on a limited number of simulation runs.

A classic approach to multi-fidelity modeling is Gaussian process (GP) emulation, namely the autoregressive formulation of \cite{kenn2000}.
However, GPs scale poorly with the high-dimensional outputs common in Earth system models \citep{rasm2006}; for MPAS-Seaice, a single HF simulation generates outputs over 4.4 million spatiotemporal points.
A common remedy is to combine data-driven dimension reduction with GP modeling, reducing the emulation task to a smaller set of input-dependent weights \citep{cressie2008, benner2015, hou2022}.
For example, principal component analysis (PCA) is often used to project outputs onto a low-dimensional basis before emulating the resulting scores with GPs \citep{higd2008}.

For spatiotemporal fields, however, PCA requires vectorizing spatiotemporal outputs, which can obscure important joint space-time structure \citep{kier2001, anan2016, ju2017, zare2018}.
Other dimension reduction approaches that avoid vectorization construct a joint space-time basis by taking the Kronecker product of spatial and temporal basis functions \citep{bili2013, kono2023}, imposing a separable dependency structure which limits their ability to represent nonseparable space-time relations.
A tensor emulator based on Higher-Order Singular Value Decomposition (HOSVD) to efficiently emulate spatiotemporal simulator outputs has also been proposed in \cite{gopalan_higher-order_2022}, but only works for single-fidelity data.

Alternative remedies that do not rely on dimension reduction include GP covariance approximations.
Full covariance GPs may employ structural simplifications such as separable kernels to remain computationally feasible with high-dimensional outputs \citep{rasm2006, kono2023, mccrimmon_multivariate_2026, conti_bayesian_2010}.
However, as noted by \cite{fricker_multivariate_2013} in the multivariate output setting, separable models can be overly restrictive, a limitation that extends to spatiotemporal outputs: assuming identical correlation structure across all spatial locations and time points is often too restrictive for complex fields.
\cite{gu_parallel_2016} utilize a parallel partial approach in which separate GPs are fit at each spatial location with shared length-scale parameters, but this formulation does not explicitly capture correlations across locations.

Beyond these structural simplifications, other covariance scalability strategies include nearest-neighbor \citep{datt2016, cheng_hierarchical_2021} and Vecchia approximations \citep{vecchia1988, colombo_new_2026}, which are highly scalable in the number of spatial locations and have achieved considerable success in modeling massive spatial and spatiotemporal data.
They are directly usable on a low-dimensional index set, and the separable-covariance emulator we benchmark against approximates its spatial covariance factor this way; the difficulty is carrying the design inputs too, as a nonseparable emulator must.
Recent work \citep{finley_efficient_2019, katzfuss_scaled_2022} demonstrates scalable GP approximations for computer experiments, but only for scalar outputs and input dimensions of $d \leq 10$ with tens of thousands of model runs, a setting that differs substantially from ours.
Moreover, as input dimension increases, Euclidean distance becomes increasingly uninformative due to distance concentration \citep{goos_when_1999, xiong_clustered_2022, binois_survey_2022}, and the number of model runs required grows sharply.
These approximations are also sensitive to the choice of ordering and conditioning structure, which is not uniquely determined \citep{guin2018}, and choosing it is harder in our setting, where each input dimension has a different length scale that must itself be estimated to define meaningful neighborhoods.

Apart from GPs, neural-network-based multi-fidelity approaches \citep{meng2021, guo2022, zhang2021, davis2025} learn flexible nonlinear representations directly from data and can capture rich, high-dimensional structure, but they are generally data-hungry, requiring large training sets to learn stable representations and avoid overfitting \citep{goodfellow2016, brigato2021, rather2024}.
This requirement can be reduced by fixing part of the representation in advance rather than learning it.
For spatiotemporal outputs, encoding the spatial coordinates and month with fixed Fourier features helps a network capture fine-scale spatial structure and the seasonal cycle while retaining a learned nonlinear representation \citep{tancik2020}.
Whether such a representation outperforms a linear basis with the number of runs available here is an empirical question, which we address in Sections~\ref{sec:sim-study} and \ref{sec:mpas-results} by benchmarking against such a network in a multi-fidelity setting.

In this paper, we build on the classical emulation approach of GPs coupled with dimension reduction, which learns efficiently from a limited number of runs while remaining computationally tractable.
We employ a Tucker decomposition basis representation that preserves space-time tensor structure while flexibly capturing nonseparable spatiotemporal dependence.
Specifically, we develop a novel MF emulator that combines Tucker decomposition with an additive discrepancy model to jointly exploit MPAS-Seaice simulations at two fidelity levels.
Tucker decomposition provides mode-wise dimension reduction while preserving the intrinsic tensor structure and capturing cross-mode interactions \citep{tuck1966}.
By explicitly modeling inter-fidelity discrepancy, the approach leverages abundant LF runs and limited HF simulations to produce accurate high-resolution predictions with uncertainty quantification.
This work is the first to emulate the spatiotemporal outputs of MPAS-Seaice.

\section{Data} \label{sec:data}

We studied sea-ice concentration output from MPAS-Seaice, simulated using E3SM version 3.0.2 \citep{e3smv302} in standalone mode for the Northern Hemisphere.
MPAS-Seaice was run on two quasi-uniform MPAS meshes, which are not defined on a regular grid, so spatial locations were indexed by a vector.
The meshes cover the Arctic Ocean and surrounding seas at 60~km ($n_s^{\LF} = 11{,}661$) and 30~km ($n_s^{\HF} = 33{,}530$) resolutions, corresponding to the LF and HF configurations.
These resolutions balance scientific validity and computational feasibility: the 30~km mesh better captures small-scale processes such as sea-ice deformation, while the 60~km mesh represents large-scale processes at lower computational cost.
Meshes coarser than 60~km failed to resolve key physical processes.

Each simulation spanned 1976--2009, with the first 23 years treated as spin-up and the remaining $n_y = 11$ years (1999--2009) used for analysis.
Model output, originally produced at 10-day intervals, was aggregated to monthly means, yielding $n_m = 12$ outputs per year.
Thus, each MPAS-Seaice run was represented as a third-order tensor with modes for spatial location, month, and year, capturing seasonal and interannual variation.

We generated ensembles of simulation runs by varying $p=6$ input parameters: the thermal conductivity of snow (\texttt{ksno}), the neutral ocean-ice drag (\texttt{dragio}), the drainage time scale of ponds (\texttt{lambda\_pond}), two parameters controlling snow grain size (\texttt{rsnw\_mlt} and \texttt{R\_snw}), and the solid fraction at the ice-ocean interface (\texttt{phi\_i\_mushy}).
These parameters were selected based on sensitivity analyses reported in \cite{urre2016}.
The LF design $\mathcal{X}^{\LF} = \{ \bx_1, \dots, \bx_{n_{x}^{\LF}} \}$, with $n_{x}^{\LF} = 89$, was constructed using a maximin Latin hypercube design.
The HF design consists of a nested subset $\mathcal{X}^{\HF} \subset \mathcal{X}^{\LF}$ of size $n_{x}^{\HF} = 21$.

For each spatial fidelity, outputs corresponding to different input settings were stacked into a fourth-order tensor: $\mathcal{Z}^{\LF} \in \mathbb{R}^{n_s^{\LF} \times n_m \times n_y \times n_{x}^{\LF}}$ and $\mathcal{Z}^{\HF} \in \mathbb{R}^{n_s^{\HF} \times n_m \times n_y \times n_{x}^{\HF}}$.
See Figures~S1 and S2 for representative temporal trends and example spatial fields of sea-ice concentration output from MPAS-Seaice, respectively.
Simulation outputs exhibit substantially higher variability during warmer months (June--September), which is precisely the period most relevant for understanding Arctic sea ice, as this is when rapid melt-season processes such as surface albedo feedbacks have the strongest influence on ice dynamics \citep{stro2018}.

\begin{figure}[!htb]
    \centering
    \includegraphics[width=0.65\linewidth]{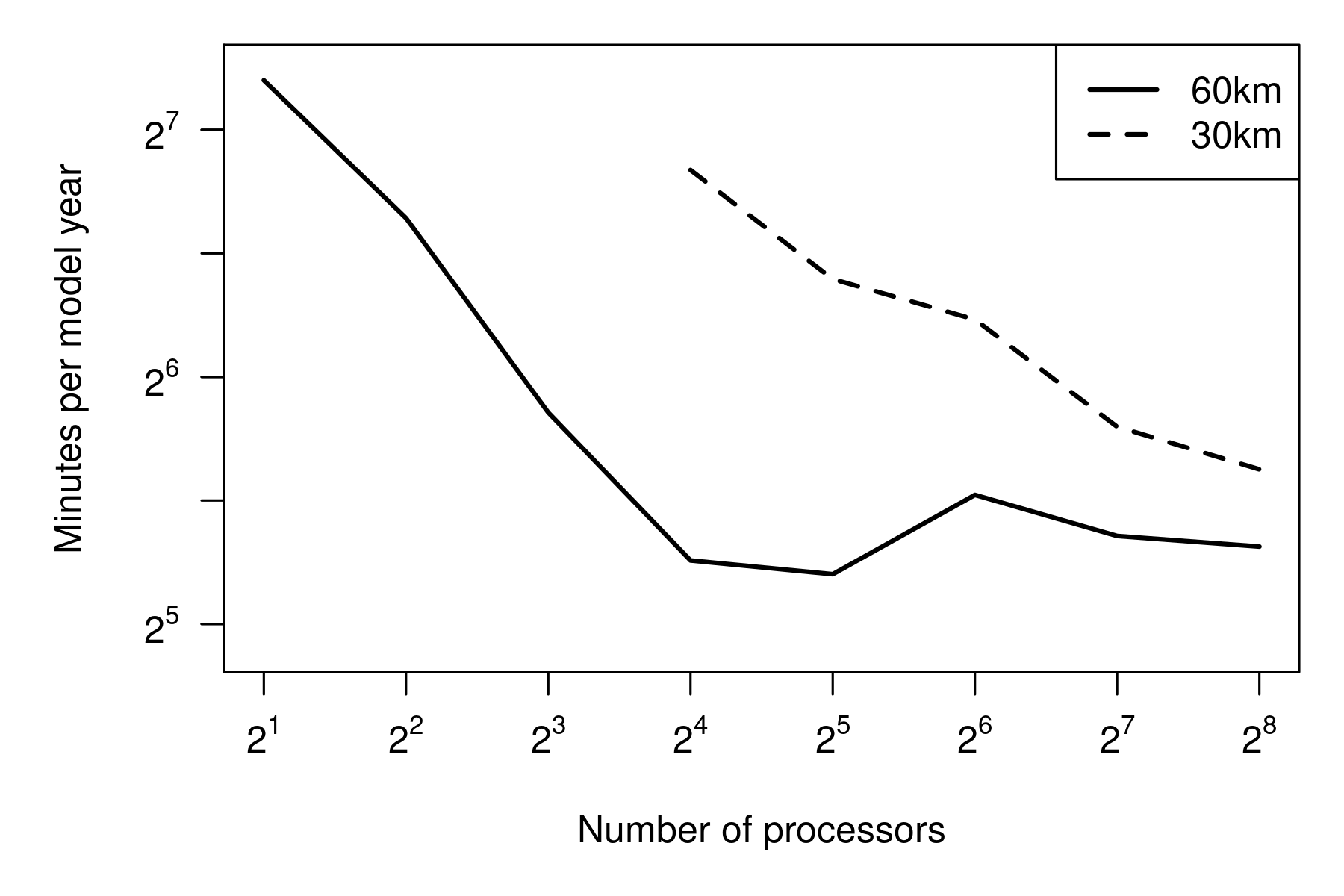}
    \caption{
        Run time comparison for 60~km (low-fidelity, LF) and 30~km (high-fidelity, HF) MPAS-Seaice simulations.
        LF simulations are up to 16 times cheaper in computational cost than the HF simulations.
    }
    \label{fig:runtime}
\end{figure}

Figure~\ref{fig:runtime} compares MPAS-Seaice runtimes for the two spatial configurations on the National Energy Research Scientific Computing Center (NERSC) Perlmutter high performance computing (HPC) system \citep{lock2021}, showing that LF simulations can reduce computational cost by up to a factor of 16 relative to HF.
This difference in computational cost motivates the development of an emulator that leverages both fidelity levels.

The MPAS-Seaice outputs are bounded in $[0,1]$, representing the proportion of each grid cell covered by sea ice.
Direct application of basis decomposition methods such as PCA or Tucker decomposition is problematic because these methods operate on unconstrained real-valued spaces and can otherwise yield invalid bounded predictions \citep{gior2007}.
To address this, we introduce a monotone transformation that maps the data to $\mathbb{R}$ while preserving the underlying spatiotemporal structure.
We use a smoothed logit-based transform, applied after clipping values to $[0.01,0.99]$, given by
\begin{equation*}
    f(x) = \logit\left[\frac{\exp(x / \epsilon) - 1}{\exp(1 / \epsilon) - 1}\right],
\end{equation*}
where a smoothing parameter $\epsilon = 10^{-3}$ pulls values near the boundaries toward the clipped limits, producing a near-linear mapping over the interior range (see Figure~S3).
Emulator predictions are mapped back to $[0,1]$ using the inverse transformation.
This approach outperformed a logit-only transform, avoiding distortions in transition zones between high and low sea-ice concentration.

\section{Methods} \label{sec:methods}

We first introduce Tucker decomposition for tensors in Section~\ref{sec:methods-tucker}.
Then, we present a framework for emulating single-fidelity tensor output in Section~\ref{sec:sf-method}, followed by the multi-fidelity tensor emulator in Section~\ref{sec:mf-method}.
Motivated by the MPAS-Seaice data, we consider simulator tensor output that is third-order (spatial location, month, and year) and, in the MF case, has two fidelities.
The methods presented can be applied to tensors of arbitrary order and, in Section~\ref{sec:mf-multi}, we discuss extension to multiple fidelities with nested designs.

\subsection{Tucker Decomposition of Model Output} \label{sec:methods-tucker}

Let $M(\cdot)$ denote a deterministic computer model mapping $\bx \in \mathbb{R}^p$ to spatiotemporal output $M(\bx) \in \mathbb{R}^{n_s \times n_m \times n_y}$, where $n_s$ is the number of spatial locations, $n_m$ is the number of temporal points per year (e.g., months), and $n_y$ is the number of years.
Combining the simulator output tensors for a set of $n_x$ design points $\mathcal{X} = \{\bx_1, \dots, \bx_{n_x}\}$, we form the single-fidelity tensor $\mathcal{Z} \in \mathbb{R}^{n_s \times n_m \times n_y \times n_x}$.

Central to our framework is to construct a set of bases from a low-dimensional representation of these outputs.
The Tucker decomposition of $\mathcal{Z}$ is
\begin{equation} \label{eq:tucker-decomp}
    \mathcal{Z} \approx \mathcal{G} \times_1 U_s \times_2 U_m \times_3 U_y \times_4 U_x,
\end{equation}
where $\mathcal{G} \in \mathbb{R}^{r_s \times r_m \times r_y \times r_x}$ is the core tensor and $U_s\in\mathbb{R}^{n_s\times r_s}$, $U_m\in\mathbb{R}^{n_m\times r_m}$, $U_y\in\mathbb{R}^{n_y\times r_y}$, and $U_x\in\mathbb{R}^{n_x\times r_x}$ are orthonormal factor matrices \citep{tuck1966, kold2009}, and $\times_k$ denotes the $k$-mode product operator for tensors.
For example, the mode-2 product $\mathcal{G} \times_2 U_m$ multiplies $\mathcal{G}$ along its second mode by $U_m$, replacing the latent month dimension $r_m$ with the observed dimension $n_m$, yielding a tensor of size $r_s \times n_m \times r_y \times r_x$ with element
\begin{equation*}
    (\mathcal{G} \times_2 U_m)[j_s, j', j_y, j_x] =
    \sum_{j_m=1}^{r_m} \mathcal{G}[j_s, j_m, j_y, j_x] \cdot U_m[j', j_m],
\end{equation*}
where $j' \in \{1, \dots, n_m\}$ indexes the observed months.
Applying all four mode products in \eqref{eq:tucker-decomp} sequentially expands $\mathcal{G}$ back to the full data dimension $n_s \times n_m \times n_y \times n_x$.
The ranks $(r_s, r_m, r_y, r_x)$ control the level of dimension reduction along each mode of $\mathcal{Z}$, with smaller ranks yielding greater compression and approximation error.
Note that $r_s\le n_s$, $r_m\le n_m$, $r_y\le n_y$, and $r_x\le n_x$, where the rank is typically substantially smaller than the original dimension.
Conceptually, Tucker decomposition is similar to singular value decomposition (SVD) for matrices and generalizes to tensors of any dimension.
For a tensor of order 2 (a matrix), SVD is a special case of Tucker decomposition with additional constraints that the core matrix is diagonal with non-negative diagonal entries.

The factor matrices have orthonormal columns and define mode-specific basis vectors for space, month, year, and the design space.
In particular, the columns of $U_s$ represent spatial patterns, $U_m$ captures seasonal (monthly) variation, and $U_y$ captures interannual variability, while $U_x$ characterizes variation across the design space.
We model month and year as separate modes because they exhibit qualitatively different behavior: the monthly mode captures the strong, recurring seasonal cycle of sea ice, while the yearly mode captures slower interannual trends and variability.
Representing them separately allows the decomposition to learn low-rank structure specific to each, rather than forcing a single temporal mode to absorb both.
The four-dimensional core tensor $\mathcal{G}$ has one index for each of the spatial, monthly, yearly, and design-space modes; an entry of large magnitude indicates that the corresponding patterns across these modes jointly contribute strongly to the simulator output.
See Supplementary Material~B for an illustration of Tucker decomposition.

Implementing Tucker decomposition requires first selecting the ranks $(r_s, r_m, r_y, r_x)$ and then computing the corresponding factor matrices and core tensor.
We choose the ranks by iteratively increasing each rank, computing the decomposition, and evaluating the percentage of Frobenius norm explained, stopping once a target percentage is reached.
For a given set of ranks, the core tensor and factor matrices are obtained using Higher-Order Orthogonal Iteration (HOOI), a block coordinate descent algorithm that yields orthonormal factor matrices and locally minimizes the approximation error \citep{dela2000b}.
See Supplementary Material~B.1 for details of the HOOI algorithm; it is computed using the \texttt{rTensor} package \citep{li2018}.

Tucker decomposition provides several advantages in our applications.
First, unlike approaches that vectorize outputs such as PCA, it does not require flattening the data, thereby preserving the multi-mode spatiotemporal structure.
Second, the factor matrices are estimated in a data-driven manner via HOOI, which learns the dominant patterns of variation along each mode directly from the model runs; the resulting factor matrices $U_s$, $U_m$, and $U_y$ correspond to interpretable spatial and temporal basis vectors, while $U_x$ captures low-dimensional variation across the design space.
Third, Tucker allows cross-mode interactions through the core tensor, which retains only those interactions supported by the data, capturing the dominant modes of spatiotemporal variation and yielding a parsimonious set of nonseparable bases.
Fourth, Tucker allows the ranks to differ across modes (i.e., $r_s \neq r_m \neq r_y \neq r_x$), which is important when the complexity of simulator output varies by mode; in contrast, canonical polyadic decomposition imposes a shared rank across all modes \citep{kold2009}.

\subsection{Single-Fidelity Tensor Emulation} \label{sec:sf-method}

We now apply Tucker decomposition to a single-fidelity setting.
Given outputs $M(\bx)$ for $\bx \in \mathcal{X}$, our objective is to construct an emulator $\eta(\bx^\star)$ that predicts $M(\bx^\star)$ at new inputs $\bx^\star \notin \mathcal{X}$, while quantifying predictive uncertainty.

\subsubsection{Emulation via Tucker Decomposition} \label{sec:sf-tuck}

For a single fidelity, we construct a tensor emulator via basis function representation with bases $\mathcal{B}_{i}$ and weights $w_{i}(\bx)$:
\begin{equation} \label{eq:basis-decomp}
    \eta(\bx) = \sum_{i = 1}^{r_x}\mathcal{B}_{i} w_{i} (\bx) + \mathcal{E}_{\eta},
\end{equation}
where $r_x$ denotes the number of basis--weight pairs.
Each $\mathcal{B}_{i}$ is a $n_s \times n_m \times n_y$ tensor capturing spatiotemporal structure, while the corresponding weights vary over the design space.
Approximation error is represented by $\bepsilon_\eta = \vect(\mathcal{E}_\eta) \sim \mathcal{N}\left(\bzero, \lambda_\eta^{-1} I\right)$ with precision $\lambda_\eta$.

The structure in \eqref{eq:tucker-decomp} provides a natural bridge to the basis representation in \eqref{eq:basis-decomp}.
Fixing the fourth-mode index $i$ of the core tensor extracts subtensor $\mathcal{G}[\, ,\, ,\, , i] \in \mathbb{R}^{r_s \times r_m \times r_y}$, which encodes the spatiotemporal interaction structure associated with the $i$-th latent design factor, $i = 1, \dots, r_x$.
Expanding this subtensor back into the original data space using the spatial and temporal factor matrices yields the $i$-th spatiotemporal basis tensor
\begin{align}\label{eq:sf-bases}
    \mathcal{B}_i
     & = \mathcal{G}[\, ,\, ,\, , i] \times_1 U_s \times_2 U_m \times_3 U_y \\
     & = \sum_{j_s = 1}^{r_s} \sum_{j_m = 1}^{r_m} \sum_{j_y = 1}^{r_y}
    \mathcal{G}[j_s,\, j_m,\, j_y,\, i] \;
    U_s[:,\, j_s] \circ U_m[:,\, j_m] \circ U_y[:,\, j_y],\label{eq:abc}
\end{align}
where $\circ$ denotes the outer product and bracket notation indexes tensor elements.
Equation~\eqref{eq:abc} makes explicit that the core tensor supplies the coefficients by which the separable space-time components $U_s[:,\, j_s] \circ U_m[:,\, j_m] \circ U_y[:,\, j_y]$ are combined.
In this sense, the core tensor plays a role analogous to the diagonal matrix of singular values in SVD, but generalized to higher-order interactions.
Note that while the bases $\mathcal{B}_i$ are constructed via a linear combination of rank-1 (i.e., separable) components, the resulting basis itself is not separable; a linear combination of rank-1 tensors is not necessarily rank-1.

The factor matrix $U_x$ holds the weights on the spatiotemporal basis tensors at the $n_x$ training design points.
Specifically, the $i$-th column of $U_x$ contains the realizations of the weight function $w_i(\bx)$ at the training designs, $\hat{w}_i(\bx_1), \dots, \hat{w}_i(\bx_{n_x})$, corresponding to the $n_x$ model runs.
Since $r_x \ll n_x$, this provides a low-dimensional representation of the simulator output across the design space; the full $n_x$ model runs are compressed into $r_x$ weight functions, each of which is subsequently modeled as a smooth GP function $w_i(\bx)$ of the input to enable prediction at new settings.

To facilitate modeling in a Bayesian hierarchical model, we express the Tucker decomposition from \eqref{eq:tucker-decomp} in matrix and vectorized forms that are convenient for statistical modeling.
Let $Z \in \mathbb{R}^{n \times n_x}$, where $n = n_s n_m n_y$, denote the matricization of $\mathcal{Z}$ obtained by stacking each tensor output $\vect(M(\bx_i))$ as a column.
Similarly, define the basis matrix $B_\eta = [\vect(\mathcal{B}_1), \dots, \vect(\mathcal{B}_{r_x})] \in \mathbb{R}^{n \times r_x}$ by vectorizing the basis tensors obtained from the Tucker decomposition.
Under this notation,
\begin{equation}\label{eq:sf-mat}
    Z \approx B_\eta U_x^\top.
\end{equation}
Vectorizing both sides of \eqref{eq:sf-mat} yields a standard linear model form.
Collect the weights as $\hat{\bw} = \vect(U_x) \in \mathbb{R}^{n_x r_x}$ and let $\bz = \vect(Z^\top) \in \mathbb{R}^{n n_x}$ order the outputs by spatiotemporal location, then by design point.
Using properties of the Kronecker product, we obtain
\begin{equation}\label{eq:sf-vec}
    \bz \approx \vect\left(U_x B_\eta^\top\right) = (B_\eta \otimes I_{n_x})\,\vect(U_x) \coloneqq B \hat{\bw},
\end{equation}
where $I_{n_x}$ is the $n_x \times n_x$ identity matrix; the ordering of $\bz$ is immaterial, as the approximation error is isotropic.
The hat on $\hat{\bw}$ denotes that these are estimates derived from the observed data, distinguishing them from the latent weight functions in \eqref{eq:basis-decomp}.
We do not place hats on the basis tensors $\mathcal{B}_{i}$ because, following common practice in reduced-rank GP emulation and PCA/basis-function emulation, we adopt a plug-in approach: the bases obtained from the Tucker decomposition are treated as fixed (empirical) quantities during subsequent inference, while uncertainty is propagated through the input-dependent weight processes.

\subsubsection{Probability Model and Posterior Inference} \label{sec:post-infer}

Given the bases and weights from Section~\ref{sec:sf-tuck}, emulation reduces to modeling the weights $w_i(\cdot)$ across the design space.
The $r_x$ columns of $U_x$ are orthonormal, so we place independent GP priors on each weight $w_i(\cdot)$.
Each GP has mean zero and a squared-exponential kernel with automatic relevance determination \citep{rasm2006}, with precision $\lambda_i$ and length scales $\rho_{i\ell}$, $\ell = 1, \dots, p$:
\begin{equation*}
    \cov(w_i(\bx),w_i(\bx')) = k_i(\bx, \bx') = \frac{1}{\lambda_i} \, \exp \left[ -\frac{1}{2} \left(
        \sum_{\ell = 1}^{p}
        \frac{(x_\ell - x'_\ell)^2}{\rho_{i\ell}}
        \right)
        \right]
    , \quad i = 1, \dots, r_x.
\end{equation*}
Let $\Sigma_{w_i}(\mathcal{X}, \mathcal{X})$ denote the covariance matrix for weight $w_i(\cdot)$ evaluated at $\mathcal{X}$.
The GP priors for all weights are combined into a block-diagonal prior with covariance
\[
    \Sigma_{w}(\mathcal{X}, \mathcal{X}) = \blockDiag[\Sigma_{w_1}(\mathcal{X}, \mathcal{X}), \dots, \Sigma_{w_{r_x}}(\mathcal{X}, \mathcal{X})].
\]
We place independent folded-normal priors on each precision $\lambda_i$, and independent log-normal priors on each length-scale parameter $\rho_{i\ell}$; these hyperparameters, along with $\lambda_\eta$, are updated sequentially via MCMC.
For brevity, we suppress the GP hyperparameters from the model displays that follow and defer details of prior specifications and the MCMC sampling scheme to Supplementary Material~E.4.

With basis functions, GP priors on the weights, and approximation error, the Bayesian hierarchical model is
\begin{equation} \label{eq:full-model}
    \begin{aligned}
        \bz \mid \bw, \lambda_{\eta} & \sim \mathcal{N} \left(B \bw, \lambda_{\eta}^{-1} I\right)                 \\
        \bw                          & \sim \mathcal{N} \left(\bzero, \Sigma_{w}(\mathcal{X}, \mathcal{X})\right) \\
        \lambda_{\eta}               & \sim \Gamma\left(a_\eta, b_\eta\right),
    \end{aligned}
\end{equation}
where $\bw$ are the latent weights and $a_\eta = 1, b_\eta = 1/2$.
Model~\eqref{eq:full-model} can be simplified by replacing $\bz$ with the observed $\hat{\bw}$ and integrating out the latent weights $\bw$ \citep{higd2008},
\begin{equation} \label{eq:model-reduce}
    \begin{aligned}
        \hat{\bw} \mid \lambda_{\eta} & \sim \mathcal{N} \left(\bzero, \lambda_{\eta}^{-1} (B^\top B)^{-1} + \Sigma_{w}(\mathcal{X}, \mathcal{X})\right) \\
        \lambda_{\eta}                & \sim \Gamma(a_\eta', b_\eta'),
    \end{aligned}
\end{equation}
where $a_\eta' = a_\eta + n_x(n - r_x) / 2$ and $b_\eta' = b_\eta + \bz^\top [I - B(B^\top B)^{-1} B^\top]\bz / 2$.
The matrix $B$ has a sparse structure that we exploit for computational efficiency; see Supplementary Materials~C.1, C.3, and E.4 for the derivation of \eqref{eq:model-reduce}, efficient calculation of $B^\top B$ and $B^\top \bz$, and MCMC details, respectively.

\subsubsection{Emulation via Posterior Prediction} \label{sec:post-pred}

Emulator $\eta$ predicts $M(\bx^\star)$ at new inputs $\bx^\star \notin \mathcal{X}$.
From the joint distribution of the observed and unobserved weights,
\begin{equation*}
    \left[
        \begin{array}{c}
            \hat{\bw} \\
            \bw^\star
        \end{array}
        \right]
    \sim
    \mathcal{N} \left(
    \bzero
    ,
    \left[
    \begin{array}{c@{\hspace{1em}}c}
    \lambda_{\eta}^{-1} (B^\top B)^{-1} + \Sigma_{w}(\mathcal{X}, \mathcal{X})
    & \Sigma_{w}(\mathcal{X}, \bx^\star) \\
    \Sigma_{w}(\bx^\star, \mathcal{X})
    & \Sigma_{w}(\bx^\star, \bx^\star)
    \end{array}
    \right]
    \right),
\end{equation*}
we derive the conditional distribution of $\bw^\star$ given $\hat{\bw}$ via standard multivariate normal results.
For each posterior predictive draw of $\bw^\star$ from this conditional distribution, the full spatiotemporal prediction is obtained by $\vect[\eta(\bx^\star)] = B_\eta \bw^\star + \bepsilon_\eta$.

\subsection{Multi-Fidelity Tensor Emulation} \label{sec:mf-method}

We extend the single-fidelity emulator of Section~\ref{sec:sf-method} to a MF setting with simulators $M^{(f)}(\bx)\in \mathbb{R}^{n_s^{(f)}\times n_m \times n_y}, f = 1, \dots, F$ of increasing fidelity in $f$, where higher fidelity corresponds to higher spatial resolution $n_s^{(f)}$.
For clarity, we describe the method for the two-fidelity case $F = 2$, writing $M^{\LF}(\cdot)$ and $M^{\HF}(\cdot)$ for the low- and high-fidelity simulators.
The extension to general $F$ is straightforward and requires no change to the model structure; we defer it to Section~\ref{sec:mf-multi}.
We do not require the spatiotemporal locations to be nested across fidelities; we do, however, assume the design points are nested, $\mathcal{X}^{\HF}\subset\mathcal{X}^{\LF}$, as is the case for our MPAS-Seaice ensemble.

Following the additive discrepancy model of \cite{kenn2000}, we represent the HF output as the interpolated LF process $\tilde{\eta}^{\LF}(\bx)$ plus a discrepancy $\delta(\bx)$ and an approximation error $\mathcal{E}_\delta$,
\begin{equation} \label{eq:additive-discrepancy}
    \eta^{\HF}(\bx) = \tilde{\eta}^{\LF}(\bx) + \delta(\bx) + \mathcal{E}_\delta, \quad \vect(\mathcal{E}_\delta) \sim \mathcal{N}\left(\bzero, \lambda_\delta^{-1} I\right).
\end{equation}
In contrast to the traditional Kennedy--O'Hagan framework, which links the fidelities using the observed LF output, we instead use the latent emulated process in order to propagate uncertainty from the LF to the HF level.
Latent-process formulations of this kind have previously been used for multi-fidelity spatial data fusion in cokriging \citep{cheng_hierarchical_2021, kono2023}.

\subsubsection{Interpolated Low-Fidelity} \label{sec:interp-lf}

We first fit a LF emulator $\eta^{\LF}$ using only the LF ensemble $\mathcal{Z}^{\LF}$, a tensor of size $n_s^{\LF}\times n_m \times n_y \times n_x^{\LF}$, following the single-fidelity approach in Section~\ref{sec:sf-method}.
This yields LF spatiotemporal bases $\mathcal{B}_i$, $i = 1, \dots, r_x$, and observed weights $\hat{\bw}^{\LF}$ at the LF design points $\mathcal{X}^{\LF}$.
To enable coupling with the HF model, each LF basis tensor $\mathcal{B}_{i}$ is spatially interpolated to the HF mesh by a weighted nearest neighbor algorithm, yielding the interpolated basis $\tilde{\mathcal{B}}_{i} \in \mathbb{R}^{n_s^{\HF} \times n_m \times n_y}$.
From these we define the matrices $\tilde{B}_\eta$ and $\tilde{B}$ analogously to \eqref{eq:sf-mat} and \eqref{eq:sf-vec}.
The interpolated LF emulator is
\begin{equation*}
    \tilde{\eta}^{\LF}(\bx) = \sum_{i=1}^{r_x} \tilde{\mathcal{B}}_{i} \, w_{i}(\bx),
\end{equation*}
which uses the same latent weight functions $w_i(\cdot)$ as the LF emulator.
Because $\mathcal{X}^{\HF} \subset \mathcal{X}^{\LF}$, the value of $\tilde{\eta}^{\LF}$ at the HF design points is obtained directly by restricting $\hat{\bw}^{\LF}$ to the HF subset, $\hat{\bw}^{\LF}\mid_{\mathcal{X}^{\HF}}$, with no prediction required.

\subsubsection{Discrepancy} \label{sec:discrep-mod}

The discrepancy term $\delta(\bx) \in \mathbb{R}^{n_s^{\HF} \times n_m \times n_y}$ corrects for systematic differences between the fidelities and exhibits structured spatiotemporal dependence; it also absorbs error introduced by the spatial interpolation of the bases.

We model the discrepancy using a separate tensor basis representation, constructed analogously to the single-fidelity emulator.
Letting $\eta^{\HF}(\bx) = M^{\HF}(\bx)$ for $\bx \in \mathcal{X}^{\HF}$ and evaluating $\tilde\eta^{\LF}$ using the restricted weights $\hat{\bw}^{\LF}\mid_{\mathcal{X}^{\HF}}$, we compute the differences $ \eta^{\HF}(\bx) - \tilde{\eta}^{\LF}(\bx)$.
These are stacked in an $n_{s}^{\HF} \times n_m \times n_y \times n_{x}^{\HF}$ tensor, to which we apply a Tucker decomposition with ranks $(r_s', r_m', r_y', r_x')$, yielding a core tensor $\mathcal{G}'$ and factor matrices $U_s', U_m', U_y', U_x'$.
In a similar fashion to Section~\ref{sec:sf-tuck}, these components produce discrepancy bases $\mathcal{D}_{i'}$ and input-dependent weights $v_{i'}(\cdot)$, $i' = 1, \dots, r_x'$.
Define matrices $D_\delta$ and $D$ from these bases as in \eqref{eq:sf-mat} and \eqref{eq:sf-vec}.
Therefore, in a similar fashion to \eqref{eq:basis-decomp}, we have
\begin{equation*}
    \delta(\bx) = \sum_{i'=1}^{r_x'} \mathcal{D}_{i'} \, v_{i'}(\bx).
\end{equation*}
The restricted weights $\hat{\bw}^{\LF}\mid_{\mathcal{X}^{\HF}}$ are only used to form the discrepancy and extract its bases $\mathcal{D}_{i'}$; they do not enter the HF model.
Consistent with the plug-in treatment of bases in Section~\ref{sec:sf-tuck}, both the interpolated LF bases and the discrepancy bases are treated as fixed empirical quantities in subsequent inference, with predictive uncertainty propagated through the input-dependent weight processes rather than through basis construction.

\subsubsection{Combined Model} \label{sec:combined-mod}

Substituting the interpolated LF term and the discrepancy term into the additive form in \eqref{eq:additive-discrepancy}, the two-fidelity emulator has the form
\begin{align*}
    \eta^{\LF}(\bx) & = \sum_{i=1}^{r_x}\mathcal{B}_i\,w_i(\bx) + \mathcal{E}_\eta,                                                                                                                               \\
    \eta^{\HF}(\bx) & = \underbrace{\sum_{i=1}^{r_x}\tilde{\mathcal{B}}_i\,w_i(\bx)}_{\tilde{\eta}^{\LF}(\bx)} + \underbrace{\sum_{i'=1}^{r'_x}\mathcal{D}_{i'}\,v_{i'}(\bx)}_{\delta(\bx)} + \mathcal{E}_\delta,
\end{align*}
where $w_i(\cdot)$ is a latent weight function shared across both fidelities.
We write $\bw_i^{\LF} = \left(w_i(\bx)\right)_{\bx\in \mathcal{X}^{\LF}}$ and $\bw_i^{\HF} = \left(w_i(\bx)\right)_{\bx\in \mathcal{X}^{\HF}}$ as the same weight process evaluated at the LF and HF design points, respectively, and let $\bw^{\LF} = \left(\bw_i^{\LF}\right)_{i=1}^{r_x}$ and $\bw^{\HF} = \left(\bw_i^{\HF}\right)_{i=1}^{r_x}$ denote the concatenation over $i$, with $\bv= \left(\left(v_{i'}(\bx)\right)_{\bx\in \mathcal{X}^{\HF}}\right)_{i'=1}^{r'_x}$ defined analogously.
As in Section~\ref{sec:post-infer}, we place independent GP priors on each $w_i(\cdot)$ and each $v_{i'}(\cdot)$.
To simplify the notation, let $\btheta = \left[\bw^{\LF\top} , \bw^{\HF\top} , \bv^\top\right]^\top$ collect all latent weights.

\subsubsection{Posterior Inference} \label{sec:mf-post-infer}

We jointly utilize LF and HF data within a single coherent Bayesian framework.
Collect the LF model runs in an ${n^{\LF} n_x^{\LF}}$-dimensional vector $\bz^{\LF} = \vect(Z^{\LF\top})$, where $Z^{\LF} \in \mathbb{R}^{n^{\LF} \times n_x^{\LF}}$ with $n^{\LF} = n_s^{\LF} n_m n_y$, is the matricization of tensor $\mathcal{Z}^{\LF}$, and similarly define the ${n^{\HF} n_x^{\HF}}$-dimensional vector $\bz^{\HF}$ collecting the HF model runs.
The joint hierarchical model is then
\begin{equation}\label{eq:joint-lk}
    \begin{aligned}
        \begin{bmatrix}
            \bz^{\LF} \\
            \bz^{\HF}
        \end{bmatrix}
        \mid \btheta, \lambda_{\eta}, \lambda_{\delta}
                     & \sim \mathcal{N} \left(
        \begin{bmatrix}
            B \bw^{\LF} \\
            \tilde{B} \bw^{\HF} + D \bv
        \end{bmatrix},
        \begin{bmatrix}
            \lambda_{\eta}^{-1} I & \bzero                  \\
            \bzero                & \lambda_{\delta}^{-1} I
        \end{bmatrix}
        \right)                                                                                                                 \\
        \btheta      & \sim \mathcal{N} \left( \bzero, \Sigma_{\theta}\right)                                                   \\
        \lambda_\eta & \sim \Gamma\left(a_\eta, b_\eta\right), \quad \lambda_\delta \sim \Gamma\left(a_\delta, b_\delta\right),
    \end{aligned}
\end{equation}
where
\begin{equation*}
    \Sigma_\theta =
    \begin{bmatrix}
        \Sigma_{w}(\mathcal{X}^{\LF}, \mathcal{X}^{\LF}) & \Sigma_{w}(\mathcal{X}^{\LF}, \mathcal{X}^{\HF}) & \bzero                                           \\
        \Sigma_{w}(\mathcal{X}^{\HF}, \mathcal{X}^{\LF}) & \Sigma_{w}(\mathcal{X}^{\HF}, \mathcal{X}^{\HF}) & \bzero                                           \\
        \bzero                                           & \bzero                                           & \Sigma_{v}(\mathcal{X}^{\HF}, \mathcal{X}^{\HF})
    \end{bmatrix}.
\end{equation*}
The cross-covariance is
\[
    \Sigma_{w}(\mathcal{X}^{\LF}, \mathcal{X}^{\HF}) = \blockDiag[\Sigma_{w_1}(\mathcal{X}^{\LF}, \mathcal{X}^{\HF}), \dots, \Sigma_{w_{r_x}}(\mathcal{X}^{\LF}, \mathcal{X}^{\HF})],
\]
where each block $\Sigma_{w_i}(\mathcal{X}^{\LF}, \mathcal{X}^{\HF}) = \cov(\bw_i^{\LF}, \bw_i^{\HF})$ is of dimension $n_x^{\LF}\times n_x^{\HF}$ and is obtained by evaluating the GP covariance kernel for the latent weight functions $w_i(\bx)$ at every pair from the subsets of LF and HF design points.

Analogous to the simplification in Section~\ref{sec:post-infer}, we derive a low-dimensional likelihood on the observed weights $\hat{\btheta} = [\hat{\bw}^{\LF\top} , \hat{\bw}^{\HF\top} ,\hat{\bv}^\top]^\top$.
The observed LF weights $\hat{\bw}^{\LF}$ come from the LF fit; the observed HF weights $\hat{\bw}^{\HF}$ and discrepancy weights $\hat{\bv}$ are not inherited from the LF fit but re-estimated from the HF data: stacking the interpolated and discrepancy bases into $K = [\tilde{B}\ \ D]$, they are obtained jointly as the least-squares projection
\begin{equation*}
    \big[\hat{\bw}^{\HF\top},\, \hat{\bv}^\top\big]^\top = (K^\top K)^{-1} K^\top \bz^{\HF},
\end{equation*}
where the first $r_x n_x^{\HF}$ entries give $\hat{\bw}^{\HF}$ and the remaining $r'_x n_x^{\HF}$ entries give $\hat{\bv}$.
Estimating the two sets of weights jointly is necessary because the discrepancy bases $D$, constructed from differences involving the interpolated LF bases, may share directions with $\tilde{B}$; the joint solution allocates this shared structure between the two components rather than to either alone.
Both $\hat{\bw}^{\LF}$ and $\hat{\bw}^{\HF}$ are thus noisy observations of the same underlying latent weight process, viewed at different design points and through different fidelity runs.
This reduces the high-dimensional likelihood in \eqref{eq:joint-lk} to
\begin{equation}\label{eq:mfm-integrated-out}
    \begin{aligned}
        \hat{\btheta}
        \mid \lambda_{\eta}, \lambda_{\delta}
         & \sim \mathcal{N}\left(
        \bzero,\Sigma_{\hat{\theta}}
        \right) , \quad
        \lambda_\eta \sim \Gamma(a_\eta', b_\eta'), \quad
        \lambda_\delta \sim \Gamma(a_\delta', b_\delta'),
    \end{aligned}
\end{equation}
where $\Sigma_{\hat{\theta}} = \Sigma_\theta + \blockDiag\left[ \lambda_{\eta}^{-1} (B^\top B)^{-1}, \; \lambda_{\delta}^{-1} (K^\top K)^{-1}\right]$.
The precision hyperparameters are updated via
\begin{equation*}
    \begin{aligned}
        a'_\eta     & = a_\eta + \frac{n_x^{\LF}(n^{\LF} - r_x)}{2},                                                                &
        b_{\eta}'   & = b_{\eta} + \frac{1}{2} \bigl(\bz^{\LF}\bigr)^\top \left( I - B (B^\top B)^{-1} B^\top \right) \bz^{\LF},      \\
        a_{\delta}' & = a_{\delta} + \frac{n_x^{\HF} [n^{\HF} - (r_x + r'_x)]}{2},                                                  &
        b_{\delta}' & = b_{\delta} + \frac{1}{2} \bigl(\bz^{\HF}\bigr)^\top \left( I - K (K^\top K)^{-1} K^\top \right) \bz^{\HF} .
    \end{aligned}
\end{equation*}
See Supplementary Material~C.2 for the derivation.
The posterior of \eqref{eq:mfm-integrated-out} is sampled via MCMC with Metropolis--Hastings updates and hyperpriors analogous to those outlined in Section~\ref{sec:post-infer}.

Restricting emulation to the weights yields a posterior that can be sampled more efficiently than that of a model without dimension reduction.
Inverting the covariance $\Sigma_{\hat{\theta}}$ scales as
\[
    \mathcal{O}\!\left[(n_x^{\LF} r_x + n_x^{\HF} (r_x + r_x'))^3\right],
\]
which is substantially cheaper than evaluating the likelihood in the unreduced setting,
\[
    \mathcal{O}\!\left[(n_x^{\LF} n^{\LF} + n_x^{\HF} n^{\HF})^3\right].
\]

\subsubsection{Posterior Prediction} \label{sec:mf-post-pred}

Unlike single-fidelity emulators, MF emulators integrate information from both LF and HF data, enhancing prediction accuracy and capturing the uncertainties in each fidelity as well as their discrepancies.
As in Section~\ref{sec:post-pred}, we derive the posterior predictive distribution for a held-out test input setting $\bx^\star$ via the conditional properties of multivariate normal distributions on the joint
\begin{equation*}
    \scalebox{0.85}{$
            \begin{aligned}
                \left[
                    \renewcommand{\arraystretch}{1.25}
                    \begin{array}{c}
                        \hat{\bw}^{\LF} \\
                        \hat{\bw}^{\HF} \\
                        \hat{\bv}       \\
                        \hline
                        \bw^{\star}     \\
                        \bv^{\star}
                    \end{array}
                    \right]
                \Big| \; \lambda_\eta, \lambda_\delta
                \sim \mathcal{N}\left(
                \bzero
                , \;
                \left[
                    \renewcommand{\arraystretch}{1.25}
                    \begin{array}{c|c}
                        \Sigma_{\hat{\theta}} &
                        \begin{array}{cc}
                            \Sigma_{w}(\mathcal{X}^{\LF}, \bx^\star) & \bzero                                   \\
                            \Sigma_{w}(\mathcal{X}^{\HF}, \bx^\star) & \bzero                                   \\
                            \bzero                                   & \Sigma_{v}(\mathcal{X}^{\HF}, \bx^\star)
                        \end{array}
                        \\ \hline
                        \begin{array}{ccc}
                            \Sigma_{w}(\bx^\star, \mathcal{X}^{\LF}) &
                            \Sigma_{w}(\bx^\star, \mathcal{X}^{\HF}) & \bzero                                            \\
                            \bzero                                   & \bzero & \Sigma_{v}(\bx^\star, \mathcal{X}^{\HF})
                        \end{array}
                                              &
                        \begin{array}{cc}
                            \Sigma_{w}(\bx^\star, \bx^\star) & \bzero                           \\
                            \bzero                           & \Sigma_{v}(\bx^\star, \bx^\star)
                        \end{array}
                    \end{array}
                    \right]
                \right).
            \end{aligned}$}
\end{equation*}
For each posterior predictive draw of $(\bw^\star, \bv^\star)$ from the resulting conditional distribution, the spatiotemporal prediction is
\begin{equation} \label{eq:mf-post-pred}
    \vect\left[{\eta}^{\HF}(\bx^{\star})\right] \mid\bz^{\LF}, \bz^{\HF}= \tilde{B}_\eta \bw^\star + D_\delta\bv^{\star} + \bepsilon_\delta.
\end{equation}

\subsubsection{Extension to Multiple Fidelities} \label{sec:mf-multi}

The two-fidelity approach generalizes to $F$ simulators, with outputs $\mathcal{Z}^{(f)} \in \mathbb{R}^{n_s^{(f)} \times n_m \times n_y \times n_x^{(f)}}$ and nested designs $\mathcal{X}^{(1)} \supset \dots \supset\mathcal{X}^{(F)}$.
For the lowest fidelity, we have
\begin{equation*}
    \eta^{(1)}(\bx) = \sum_{i=1}^{r_x}\mathcal{B}_i w_i(\bx) + \mathcal{E}_\eta.
\end{equation*}
For $f=2,\dots,F$, we follow the latent variable autoregressive form
\begin{equation*}
    \eta^{(f)}(\bx) = \tilde{\eta}^{(f-1)}(\bx) + \delta^{(f)}(\bx) + \mathcal{E}_\delta^{(f)},
\end{equation*}
where $\tilde{\eta}^{(f-1)}(\bx)$ is the $(f-1)$-fidelity emulator interpolated to the $(f)$-fidelity spatial resolution, and $\delta^{(f)}(\bx)$ is the discrepancy capturing systematic differences between fidelities $(f-1)$ and $(f)$ with bases $\mathcal{D}^{(f)}_{i'}$ and weights $v^{(f)}_{i'}(\bx)$, ${i'} = 1, \dots, {r_x'}^{(f)}$.
As in the two-fidelity case, the weight functions $w_i(\cdot)$ are shared across all fidelity levels and enter each $\tilde{\eta}^{(f-1)}$ through the interpolated bases, while each discrepancy $\delta^{(f)}$ introduces its own weights $v_{i'}^{(f)}(\cdot)$.
The highest-fidelity emulator is thus the lowest-fidelity emulator $\eta^{(1)}$ plus the discrepancy terms $\delta^{(2)}, \dots, \delta^{(F)}$, each carried to the highest spatial resolution through its interpolated bases.
This formulation yields a Bayesian hierarchical model analogous to \eqref{eq:joint-lk} that can be explored via MCMC.

\section{Simulation Study} \label{sec:sim-study}

We validated our method on synthetic data with a known analytical form, allowing direct comparison of emulator predictions to the true outputs.
The deterministic simulator, adapted from the Shubert function \citep{surj2013}, maps $\bx = [x_1, x_2, x_3]^\top \in [0,1]^3$ to spatiotemporal fields over spatial coordinates $\bs = (s_1, s_2)$ with monthly outputs over a five-year period; a second deterministic function generated an additive discrepancy that produced the HF output.
We included $x_3$ as a dummy input with no effect on the output to assess how the emulator handles irrelevant inputs.
Simulations used $25 \times 25$ (LF) and $50 \times 50$ (HF) spatial grids, with $n_x^{\LF} = 100$ and $n_x^{\HF} = 10$ design points.
As the design changes, the simulator's spatial pattern oscillates rapidly and unevenly across space, requiring $r_x = 11$ LF and $r_x' = 8$ discrepancy basis--weight pairs to retain 95\% variance.
Additional details regarding the design rationale, functional forms, and full rank selection are given in Supplementary Material~D.

We compared seven emulators.
Four utilized basis decomposition: LF-TEN and HF-TEN are single-fidelity tensor emulators (Section~\ref{sec:sf-method}) trained on LF and HF data, respectively, with LF-TEN predictions interpolated to the HF grid.
MF-TEN is the multi-fidelity tensor emulator described in Section~\ref{sec:mf-method}.
MF-PCA is a multi-fidelity emulator that replaces Tucker decomposition with PCA applied to vectorized spatiotemporal outputs, with the number of components chosen to explain the same percentage of variance as MF-TEN, extending \cite{higd2008}.
The remaining three methods avoid basis decomposition.
MF-SEP is a multi-fidelity cokriging approach with a separable covariance structure: $\Sigma = \Sigma_s \otimes \Sigma_m \otimes \Sigma_y \otimes \Sigma_x$ \citep[e.g.,][]{mccrimmon_multivariate_2026, conti_bayesian_2010}.
MF-IDP is a baseline that applies the traditional \cite{kenn2000} multi-fidelity approach to each spatiotemporal location independently.
Lastly, MF-FFN is a multi-fidelity neural network with a fixed-feature expansion, a deterministic Fourier feature bank on the spatial coordinates, included as a nonlinear benchmark for the linear Tucker representation \citep{tancik2020}.
See Supplementary Material~E.1 for additional details regarding the implementation of these methods.

Predictions were evaluated at 100 held-out test inputs $\bx^\star$ against the corresponding HF outputs, using four metrics.
Three are marginal, computed separately at each spatiotemporal location and then averaged over locations: absolute error (AE), 95\% predictive interval coverage, and the continuous ranked probability score \citep[CRPS;][]{gnei2007}.
Because marginal metrics score each location in isolation, they cannot detect whether a method's predictive distribution captures dependence across locations and times.
We therefore also report the spatiotemporal variogram score \citep{sche2015}, which assesses this dependence by measuring how well the emulator reproduces the variogram of the simulator output (smaller is better).
See Supplementary Material~E.2 for additional details on the metrics and their computation.

\begin{figure}[!htb]
    \centering
    \includegraphics[width=0.9\linewidth]{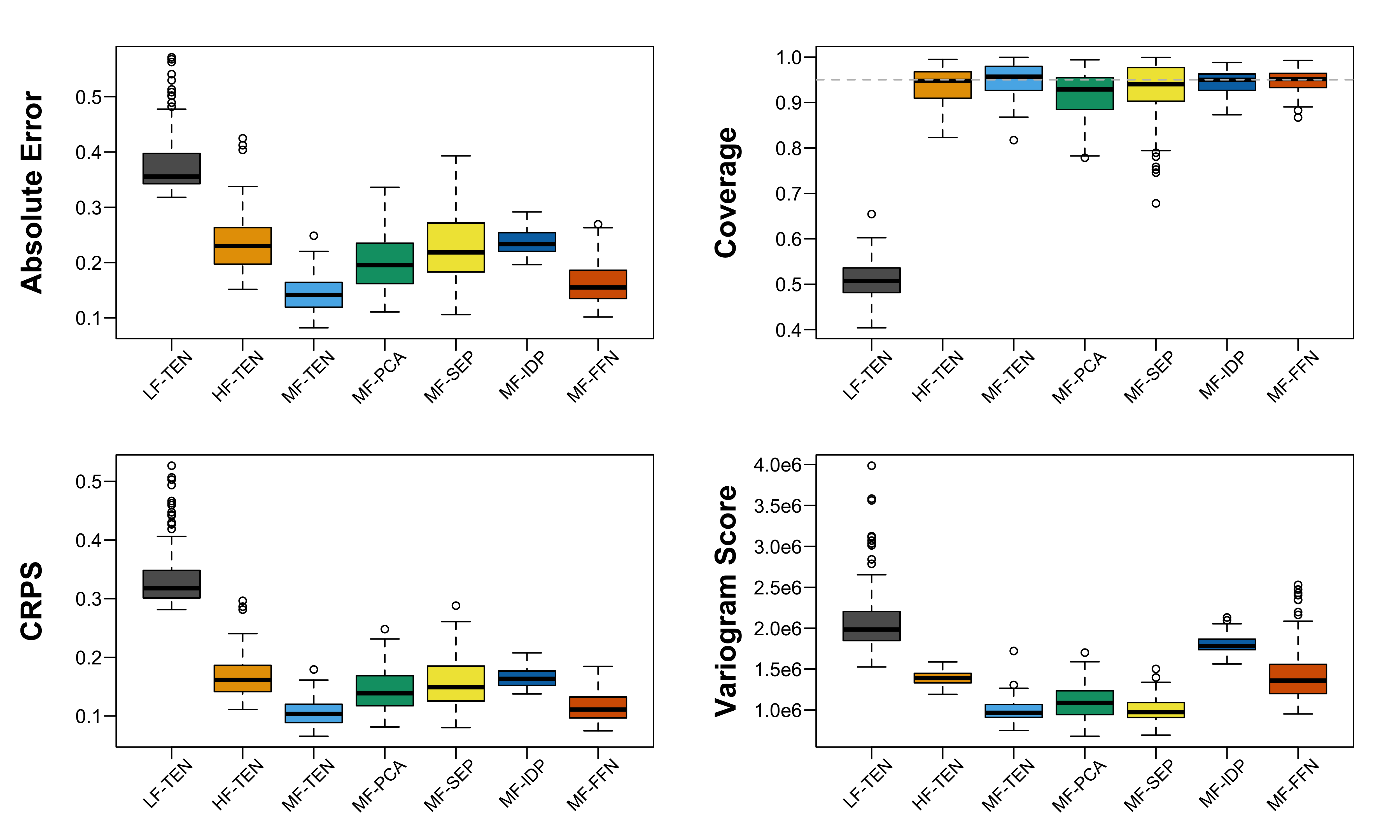}
    \caption{
        Simulation study results: box plots of absolute error, 95\% predictive interval coverage (dashed line at nominal level), continuous ranked probability score (CRPS), and spatiotemporal variogram score across 100 held-out test inputs.
        Absolute error, coverage, and CRPS are averaged over the spatiotemporal domain for each test input; the variogram score is computed jointly over that domain.
        Lower values are better for all metrics except coverage, which is best near the nominal level.
    }
    \label{fig:sim-overall}
\end{figure}

Figure~\ref{fig:sim-overall} shows that MF-TEN attains the lowest median AE and CRPS with near-nominal coverage, and a variogram score matched only by MF-SEP.
LF-TEN ignores the discrepancy and so is biased and severely under-covers, while HF-TEN, trained on only ten runs, cannot recover the input dependence that MF-TEN borrows from the LF ensemble.
MF-PCA uses the same runs and variance target as MF-TEN, so its higher AE and CRPS reflect the vectorized basis, which cannot reuse spatial patterns across months and years.
MF-SEP imposes no rank truncation but assumes a separable covariance, and it is less accurate than MF-TEN, suggesting that the information discarded by truncation matters less than the flexibility that separability gives up.
MF-IDP and MF-FFN do not explicitly model dependence across locations, so their predictive draws carry little cross-location dependence, which the variogram score penalizes.
Nonetheless, MF-FFN comes close to MF-TEN's pointwise accuracy, suggesting that a nonlinear representation offers little gain here over the linear Tucker representation, which additionally captures joint spatiotemporal dependence and provides an interpretable low-rank structure.

The MF-TEN emulator also provides input sensitivity estimates through the posterior length scales, where smaller values indicate higher sensitivity.
It correctly identifies the dummy input $x_3$ as inactive, assigning it larger length scales relative to $x_1$ and $x_2$; full results are reported in Supplementary Material~E.5.

\section{MPAS-Seaice Emulation} \label{sec:mpas-results}

We evaluated model performance using the MPAS-Seaice data; see Section~\ref{sec:data} for details.
For each of the $n_{x}^{\HF} = 21$ inputs with both LF and HF runs, we conducted leave-one-out cross-validation (LOO-CV), training seven emulators as in Section~\ref{sec:sim-study}, and making predictions for the output from the held-out design point.
Performance was evaluated using the same metrics: AE, 95\% coverage, CRPS, and variogram score.
For the MCMC-based emulators, four chains were run from different initializations, each with 10,000 iterations and a burn-in of 5,000.
Convergence was assessed using Geweke Z-scores for within-chain mixing and Gelman--Rubin diagnostics for between-chain mixing, with full MCMC details in Supplementary Material~E.4.

We determined the Tucker ranks for both the LF and HF tensors by retaining 95\% of the variance in each decomposition, resulting in $(r_s, r_m, r_y, r_x) = (79, 7, 6, 4)$.
For the discrepancy, we targeted 80\% explained variance, yielding $(r_s', r_m', r_y', r_x') = (82, 7, 7, 4)$; a higher target would have required hundreds of spatial bases.

\begin{figure}
    \centering
    \includegraphics[width=0.7\linewidth]{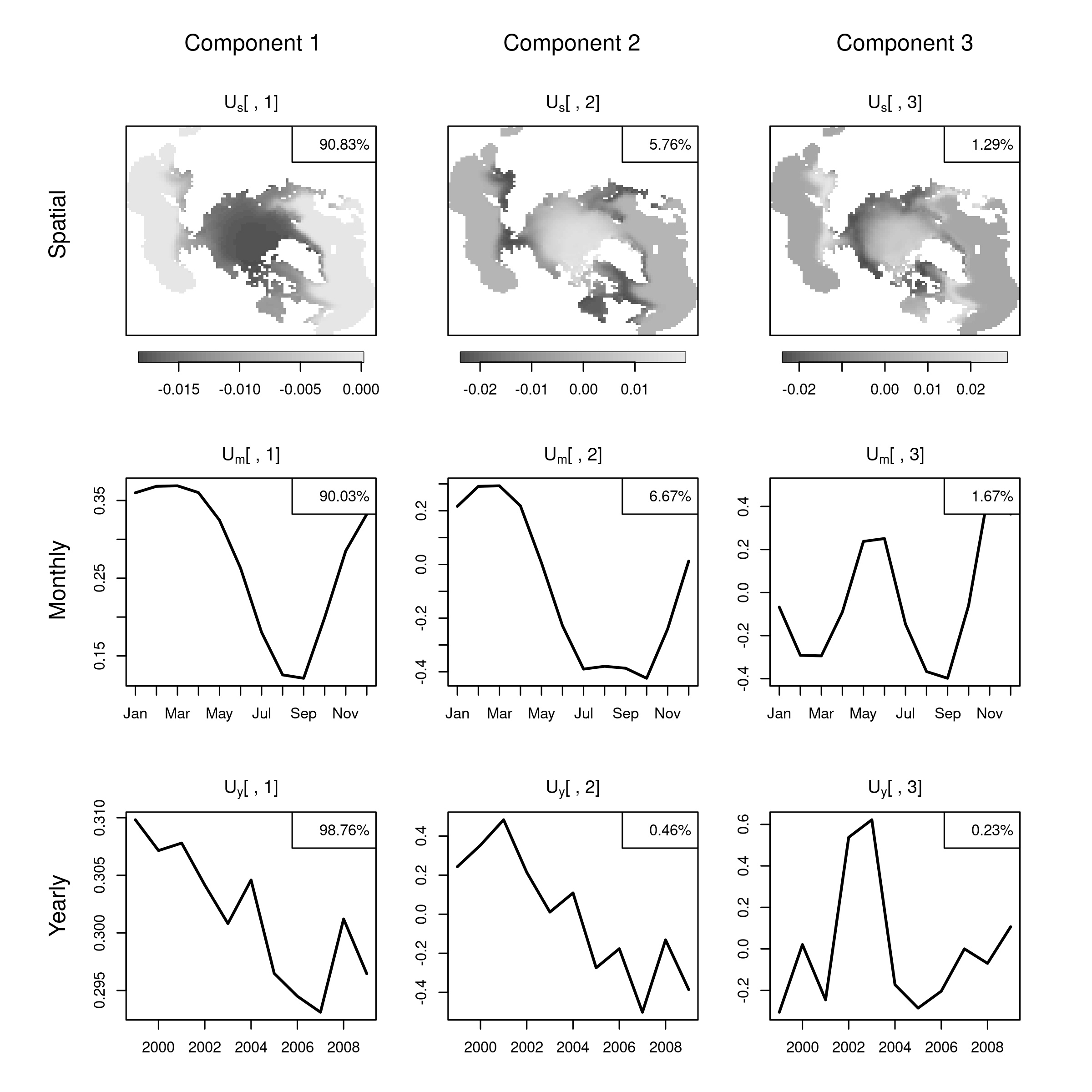}
    \caption{
        First three spatial, monthly, and yearly components from the Tucker decomposition factor matrices of low-fidelity MPAS-Seaice data.
        The percentage of variance explained is shown for each mode in the top right, with the leading component in every mode explaining over 90\% of the variance.
    }
    \label{fig:sea-ice-bases}
\end{figure}

An advantage of the Tucker decomposition is interpretability.
The decomposition produces factor matrices whose columns are interpretable mode-specific basis patterns, which we can inspect for both the LF data and the discrepancy term.
The relative percentage of variance explained by each basis is computed as the ratio of its squared core-tensor entries to the total \citep{sidi2017}.
Writing this in the LF notation, the spatial basis $j_s$ explains
\begin{equation*}
    \text{propVar}(j_s) = \frac{\sum_{j_m, j_y, j_x} \mathcal{G}[j_s, j_m, j_y, j_x]^2}
    {\sum_{j_s', j_m, j_y, j_x} \mathcal{G}[j_s', j_m, j_y, j_x]^2},
\end{equation*}
where the denominator is the total sum of squared core entries, $\lVert\mathcal{G}\rVert_F^2$.
Figure~\ref{fig:sea-ice-bases} visualizes the first three columns of the factor matrices, along with the relative percentage of variance each explains.
The leading spatial pattern represents typical ice transitions in colder months, whereas the second and third capture transitions more common in warmer months.
The first two monthly components capture the seasonal cycle, and the first two yearly components show a general decline in sea ice from 1999--2009.
The third temporal components indicate finer-scale variation.
In each mode, the leading component explains over 90\% of the variance.
Beyond this aggregated view, we can further examine how each spatial basis contributes over time through a relative importance measure, given in Supplementary Material~B.2.

\begin{figure}
    \centering
    \includegraphics[width=0.9\linewidth]{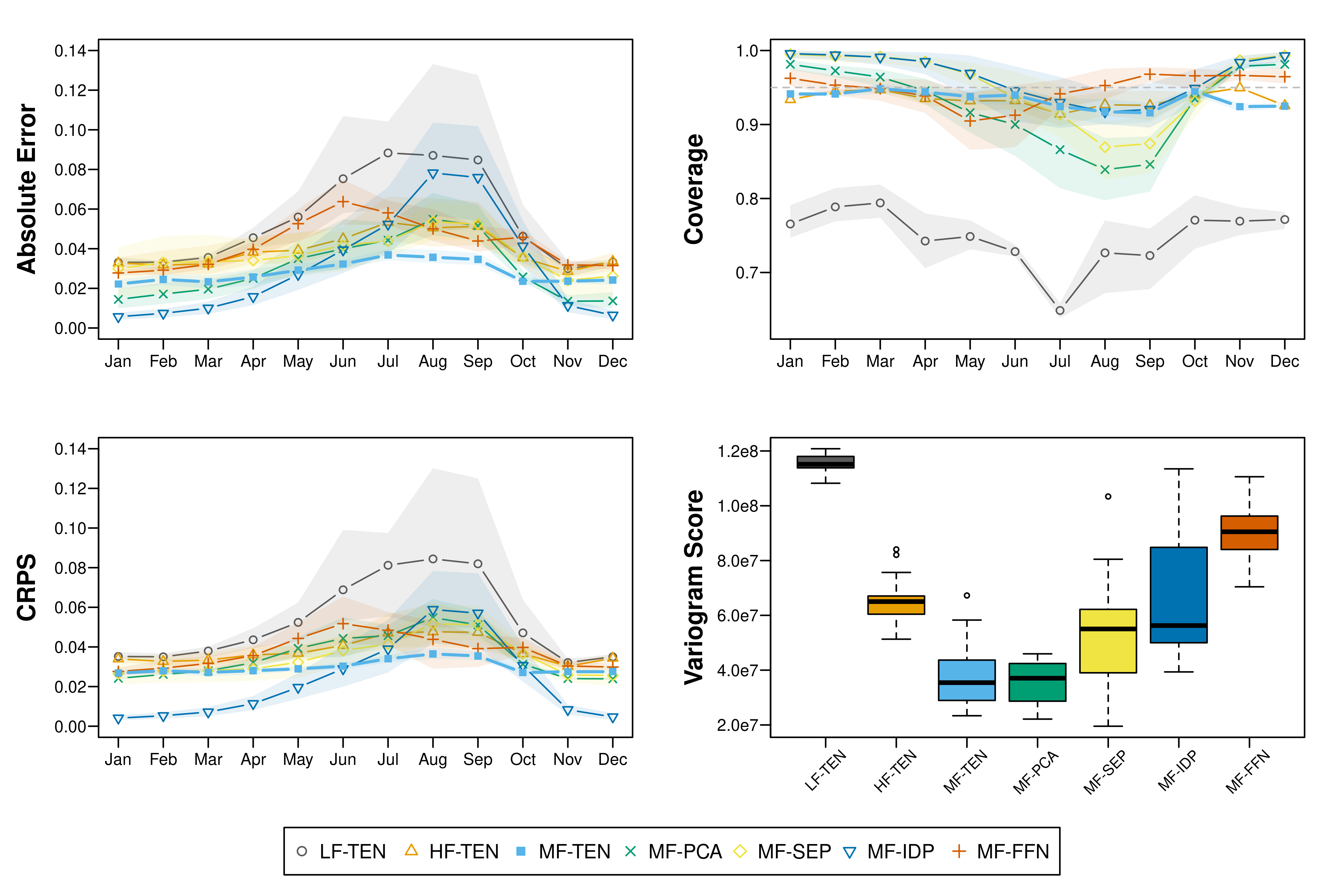}
    \caption{
        Monthly performance of emulators on MPAS-Seaice data, averaged over spatial locations and years within each LOO-CV fold.
        Lines show the mean absolute error, coverage, and CRPS, with bands giving the 2.5\%--97.5\% quantile range across folds and the dashed line marking the 95\% nominal coverage target.
        The variogram score, computed jointly over the full spatiotemporal domain per fold, is instead summarized as a box plot by method.
    }
    \label{fig:sea-ice-month}
\end{figure}

Figure~\ref{fig:sea-ice-month} summarizes the monthly performance of the emulators.
All emulators find June--September hardest, which is also the period of greatest scientific interest (Section~\ref{sec:data}), and MF-TEN has the lowest AE in each of these months; in August its predictions are on average 1.5 percentage points closer to the true sea-ice concentration than those of the next-best emulator.
Because MF-SEP's separable covariance gives every location the same predictive variance (Supplementary Material~E.1), it over-covers where ice is static (winter) and under-covers where it varies (late summer).
MF-FFN is the closest competitor in August and September and, being conformally calibrated, has better late-summer coverage, but its AE exceeds MF-TEN's in every month and its variogram score is the second-worst, after LF-TEN's.

\begin{figure}
    \centering
    \includegraphics[width=0.9\linewidth]{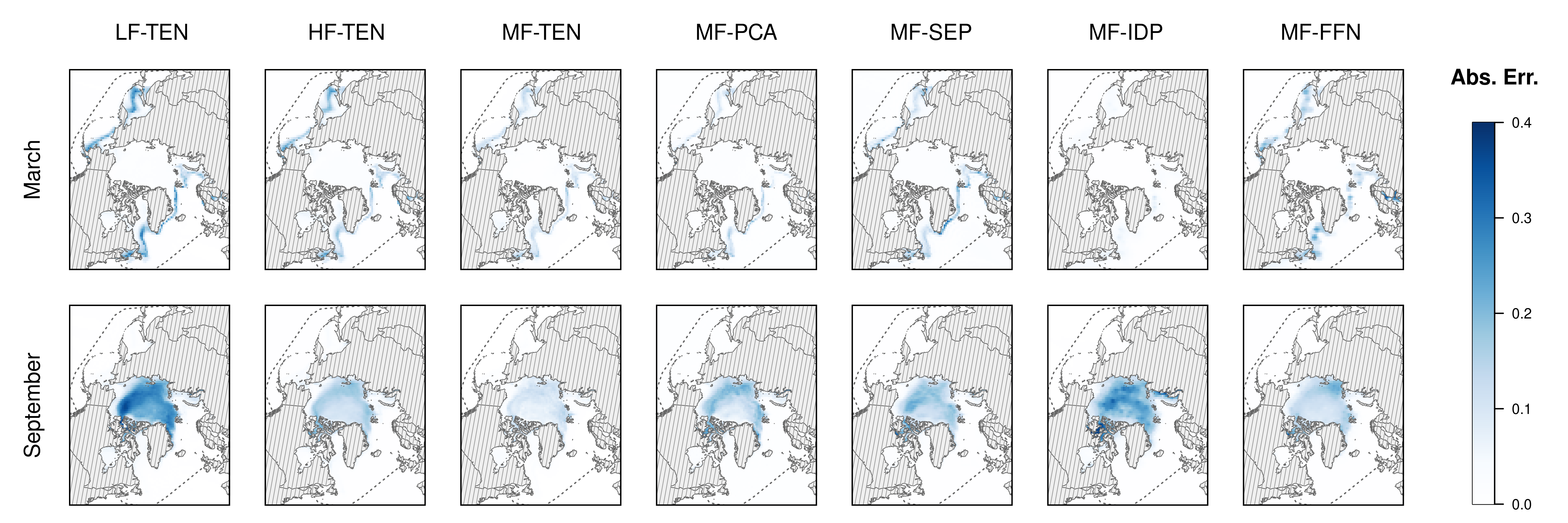}
    \includegraphics[width=0.9\linewidth]{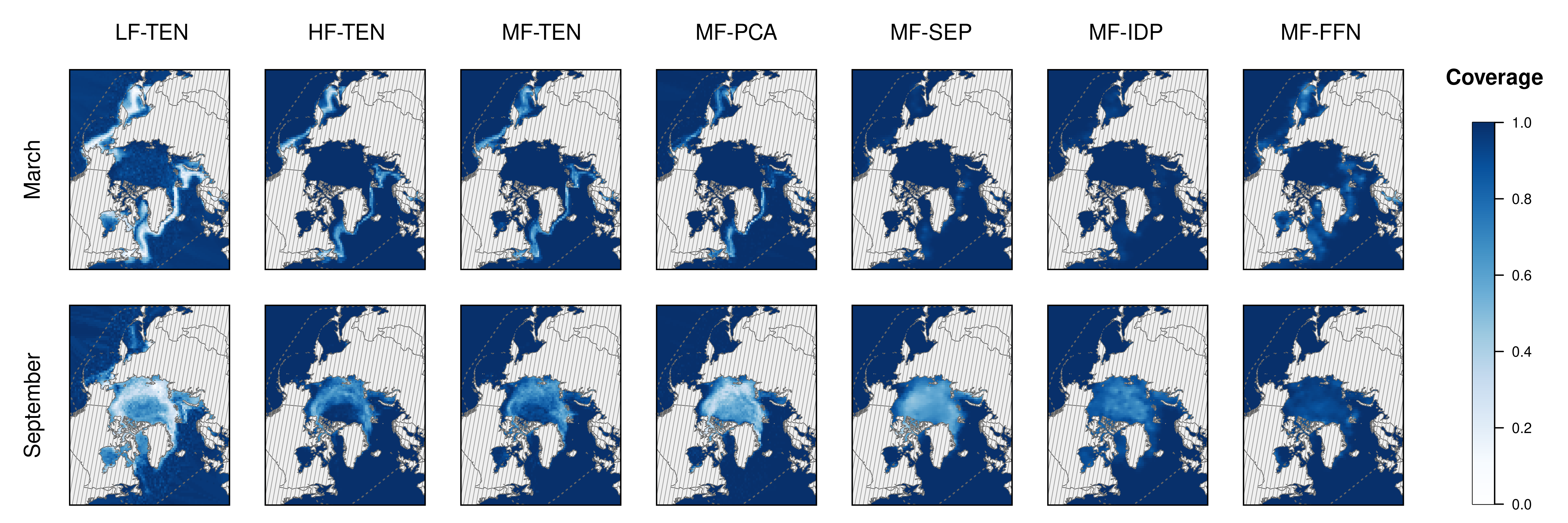}
    \caption{
        Spatial performance comparison of emulators on MPAS-Seaice data for March and September, showing absolute error (top) and 95\% predictive interval coverage (bottom) averaged over LOO-CV and year at each spatial location.
        The MPAS-Seaice domain is outlined with a dotted line.
        March exhibits higher variability in thin transitional ice bands, September across the Central Arctic and surrounding seas.
    }
    \label{fig:sea-ice-spatial}
\end{figure}

Figure~\ref{fig:sea-ice-spatial} shows that prediction is more difficult when ice concentration is more variable (see Figure~S4 for CRPS).
In September, ice concentration variability spans the Central Arctic; its structure is complex but varies less sharply across runs, so the shared spatial bases capture it and MF-TEN has the lowest AE there.
In March, ice concentration is uniform over a large portion of the Arctic, with variable concentration in a thin ice edge whose position shifts across runs.
Resolving this edge would take MF-TEN many extra bases that each add little explained variance, so in March MF-IDP (fit per location) and MF-PCA (basis not shared across months; see below) attain lower AE.

MF-IDP fits independent models at each spatiotemporal location, so it captures no correlation across space or time and its accuracy tracks local variability.
Where sea-ice concentration varies widely across design points, its AE is substantially higher than MF-TEN's; where it is constant, as at the many locations fixed at 0 or 100\% ice cover in colder months, its leave-one-out predictions are accurate by construction, with uncertainty collapsing toward zero, rather than by inference.
Fitting each location separately also lets it follow the shifting March ice edge.
MF-PCA, whose variogram score is comparable to MF-TEN's, instead lacks a spatial basis shared across months, so it can devote components to single-month features such as the cold-season ice edge, which helps in winter but forfeits the cross-month sharing that benefits MF-TEN in the warm season.

Motivated by differences between high- and low-variability regions, we explored a regime-based emulator dividing the data into four regimes: within the cold and warm months separately, two spatial regimes are defined by thresholding the variance computed across design points at each location.
Low-variance regions required fewer ranks ($r_x=2$) than high-variance regions ($r_x=6$) to explain the same 95\% of variance.
Figure~S5 shows that the regime-based emulator achieves slightly lower AE than the global MF-TEN, with a more notable reduction in prediction uncertainty.
A limitation is that these regime boundaries come from ad hoc thresholding, so nearby locations on either side can show markedly different prediction uncertainty.
Avoiding these hard boundaries with an adaptive spatiotemporal mixture of regime-specific Tucker emulators, while retaining the low-dimensional basis's efficiency, remains a promising direction for future work.

Average posterior length scales for each input variable and weight in MF-TEN, computed using the full data rather than LOO-CV, are reported in Table~\ref{tbl:sea-ice-lengthscales}.
Across all input variables, thermal conductivity of snow (\texttt{ksno}) has the smallest length scales in both the LF and discrepancy components, so MPAS-Seaice is most sensitive to this parameter.
In contrast, solid fraction at the ice-ocean interface (\texttt{phi\_i\_mushy}) has some of the largest length scales, so the response varies only weakly with it.

\begin{table}
    \caption{
        Posterior mean length scales of the LF and discrepancy terms for each weight and input variable in the MF-TEN emulator applied to MPAS-Seaice data.
        See Section~\ref{sec:data} for a description of the six input parameters.
    }
    \label{tbl:sea-ice-lengthscales}
    \centering
    \renewcommand{\arraystretch}{1.15}
    \begin{tabular}{lcccccccc}
        \toprule
                               & \multicolumn{4}{c}{\textbf{Low-Fidelity}}
                               & \multicolumn{4}{c}{\textbf{Discrepancy}}                                                             \\
        \cmidrule(lr){2-5}\cmidrule(lr){6-9}
                               & {$w_1$}                                   & {$w_2$} & {$w_3$} & {$w_4$} &
        {$v_1$}                & {$v_2$}                                   & {$v_3$} & {$v_4$}                                        \\
        \midrule
        \texttt{ksno}          & 1.22                                      & 0.47    & 0.32    & 0.39    & 2.26  & 0.32 & 0.97 & 0.34 \\
        \texttt{dragio}        & 4.12                                      & 1.40    & 0.54    & 1.28    & 4.52  & 6.03 & 1.06 & 2.48 \\
        \texttt{lambda\_pond}  & 7.47                                      & 1.21    & 2.74    & 1.59    & 5.18  & 1.57 & 8.11 & 4.17 \\
        \texttt{rsnw\_mlt}     & 6.98                                      & 2.35    & 6.94    & 0.92    & 4.19  & 1.83 & 5.29 & 2.24 \\
        \texttt{R\_snw}        & 1.98                                      & 1.89    & 4.95    & 1.48    & 12.68 & 2.01 & 3.47 & 1.99 \\
        \texttt{phi\_i\_mushy} & 6.31                                      & 7.09    & 2.93    & 5.83    & 6.63  & 5.06 & 3.63 & 8.76 \\
        \bottomrule
    \end{tabular}
\end{table}

To illustrate computational savings of MF-TEN, the reduced covariance inversion scales as
\[
    \mathcal{O}\!\left[(n_x^{\LF} r_x + n_x^{\HF} (r_x + r_x'))^3\right] = \mathcal{O}\!(524^3),
\]
whereas the unreduced setting would require
\[
    \mathcal{O}\!\left[(n_x^{\LF} n^{\LF} + n_x^{\HF} n^{\HF})^3\right] \approx \mathcal{O}\![(2.3\times 10^8)^3].
\]
In terms of wall-clock time per LOO-CV training, MF-TEN required 176 minutes, compared to 52 minutes for LF-TEN, 83 minutes for HF-TEN, 178 minutes for MF-PCA, 30 minutes for MF-SEP, 210 minutes for MF-IDP, and 21 minutes for MF-FFN.
Although MF-TEN is slower to train than most alternatives, prediction at a new input is fast, so it suits downstream tasks such as calibration, and its gains in accuracy and uncertainty quantification justify the added cost.
Because MPAS-Seaice runs on an HPC system, direct runtime comparisons are not meaningful; for scale, the time to fit MF-TEN would allow roughly 1--2 model years of HF MPAS-Seaice simulation.

\section{Discussion} \label{sec:discussion}

We propose a multi-fidelity tensor (MF-TEN) emulation framework for large-scale Earth system models that combines the computational efficiency of LF simulations with the accuracy of HF simulations.
Applied to MPAS-Seaice, MF-TEN attains the lowest AE of all emulators during the summer melt season, with near-nominal coverage and a variogram score at least as low, and its advantage is largest across the broad September transition zone in the Central Arctic.
MF-FFN comes closest, nearly matching MF-TEN in the simulation study and approaching it in late summer on MPAS-Seaice, but its substantially higher variogram score shows that it reproduces the spatiotemporal structure of the output less faithfully.
MF-TEN's advantage during warmer months is especially relevant scientifically, since this is when sea-ice concentration changes most rapidly and accurate, uncertainty-aware predictions are valuable to practitioners \citep{stro2018}.

The components of MF-TEN play complementary roles: the LF emulator exploits the larger LF ensemble to capture broad spatiotemporal structure, while the discrepancy term, motivated by systematic differences between the 30~km and 60~km meshes, such as sea-ice deformation, corrects for the resolution gap using the smaller HF ensemble.
Data-adaptive Tucker ranks accommodate the differing variability across space, month, year, and input, and the resulting Tucker bases capture nonseparable interactions.

Tucker decomposition avoids two related but distinct forms of separability.
The first is separability of the bases, constructed as $U_s \otimes U_m \otimes U_y \otimes U_x$ with $U_a \in \mathbb{R}^{n_a \times r_a}$ for each mode $a \in \{s, m, y, x\}$; this offers computational benefits \citep{bili2013}, but the number of bases $r_s r_m r_y r_x$ grows multiplicatively, and each $U_a$ is constructed independently per mode, omitting the cross-mode interactions that Tucker's bases retain (Section~\ref{sec:methods-tucker}).
The second is separability of the covariance, $\Sigma = \Sigma_s \otimes \Sigma_m \otimes \Sigma_y \otimes \Sigma_x$, pursued in the MF-SEP method \citep{conti_bayesian_2010}: under this structure, $\cov[\eta(\bs,m,y,\bx^\star), \eta(\bs',m',y',\bx) \mid \eta(\bs,m,y,\bx)] = 0$ for $(\bs',m',y') \neq (\bs,m,y)$, so information from a new input $\bx^\star$ is borrowed only through runs at the same spatiotemporal location, eliminating cross-location and cross-time information sharing \citep{fricker_multivariate_2013}.

The main limitation is that MF-TEN is linear and its bases are fixed before the weight processes are fit, so structure the bases cannot express is absorbed into the approximation error rather than predicted, and no flexibility in the GPs on the weights recovers it.
On both problems studied here, MF-TEN matched or exceeded the accuracy of a nonlinear benchmark (MF-FFN) while better capturing spatiotemporal dependence, so the added flexibility was not needed.
Nonlinear approaches such as MF-FFN may be preferable for outputs whose spatial patterns shift sharply with the inputs, and we recommend checking the variance retained at the selected ranks before applying MF-TEN elsewhere.

While our analysis focused on fourth-order tensors (space, month, year, and design input), the framework generalizes to tensors of arbitrary order.
We considered a third-order representation that combines month and year into a single temporal mode, as well as the MF-PCA formulation that collapses space, month, and year into one mode; both exhibited higher AE and CRPS than the fourth-order method on MPAS-Seaice during warmer months (see Figure~S6 for the third-order case).
Collapsing modes prevents the model from sharing spatial basis functions across months, so components specialize in individual months' features and the recurring seasonal cycle is no longer separated from interannual variability.

Our MF-TEN emulator can also be integrated into adaptive experimental design strategies, leveraging its fast, uncertainty-aware predictions to efficiently determine both where to sample next and which fidelity level to use \citep{song2019, pell2023, garb2024}.
While we focused on spatial resolution to distinguish fidelities, the method generalizes to other hierarchies such as differences in numerical precision \citep{harv2016} or the underlying complexity of simulation physics \citep{mahe2019}.
Advances in tensor decomposition could further enhance flexibility and scalability, including methods allowing more general spatial or temporal arrangements within a fidelity \citep{tang2025}, functional tensor decompositions for extrapolation to new spatiotemporal locations \citep{han2023}, and generalized CP decompositions using non-$\ell_2$ losses for robustness to outliers and non-Gaussian residuals \citep{hong2020}.
Incorporating uncertainty in the estimated basis and rank through fully Bayesian formulations \citep{zhao2015, stolf2025} remains important, though costly for high-dimensional outputs; as a practical alternative, we assessed the stability of the Tucker bases by re-estimating them across a bootstrap over the design points and found the resulting bases to be stable (Supplementary Material~F).

\newpage

\section*{Significance}

Detailed Earth-system simulations are too costly to run many times, so we introduce a method that combines many cheap, coarse simulations with a few expensive, fine ones to predict what the fine simulation would show, with a measure of confidence.
At its center is a tensor decomposition: the output forms a four-way array over location, month, year, and model setting, and a Tucker decomposition extracts a small set of shared spatial patterns and seasonal cycles along with their interactions.
Each simulation is then summarized by how strongly those patterns combine, rather than fitted at every location separately.
Applied to Arctic sea ice, its gains are largest where ice forms and melts, the conditions hardest to predict and most important scientifically.

\section*{Acknowledgments}
E3SM v3.0.2 was obtained from the E3SM project, sponsored by the U.S.\ DOE, Office of Science, Office of Biological and Environmental Research Earth Systems Model Development Program area of Earth and Environmental System Modeling.

\section*{Funding}
This material is supported by the U.S.\ Department of Energy, Office of Science, Office of Biological and Environmental Research, under Award Number DE-SC0023366.
This research used resources of the National Energy Research Scientific Computing Center (NERSC), a DOE User Facility using NERSC award BER-ERCAP-0032240.

\section*{Supplementary Materials}
The Supplementary Materials contain additional figures, derivations, and supporting results.
These materials, along with code are available at \url{https://github.com/tjcontant/mf_tensor}.

\bibliographystyle{plainnat}
\bibliography{bibliography}

\end{document}